\documentclass[aps,prb,twocolumn,floats,showpacs]{revtex4}
\usepackage{epsfig}
\usepackage{color}
\usepackage{bm}
\usepackage{latexsym}

\begin{document}
\newcommand{\hide}[1]{}
\newcommand{\tbox}[1]{\mbox{\tiny #1}}
\newcommand{\half}{\mbox{\small $\frac{1}{2}$}}
\newcommand{\sinc}{\mbox{sinc}}
\newcommand{\const}{\mbox{const}}
\newcommand{\trc}{\mbox{trace}}
\newcommand{\intt}{\int\!\!\!\!\int }
\newcommand{\ointt}{\int\!\!\!\!\int\!\!\!\!\!\circ\ }
\newcommand{\eexp}{\mbox{e}^}
\newcommand{\bra}{\left\langle}
\newcommand{\ket}{\right\rangle}
\newcommand{\EPS} {\mbox{\LARGE $\epsilon$}}
\newcommand{\ar}{\mathsf r}
\newcommand{\im}{\mbox{Im}}
\newcommand{\re}{\mbox{Re}}
\newcommand{\bmsf}[1]{\bm{\mathsf{#1}}}
\newcommand{\mpg}[2][1.0\hsize]{\begin{minipage}[b]{#1}{#2}\end{minipage}}

\title{Scattering and
transport statistics at the metal-insulator transition: A numerical
study of the power-law banded random matrix model}

\author{J. A. M\'endez-Berm\'udez}
\affiliation{Instituto de F\'{\i}sica, Benem\'erita Universidad Aut\'onoma de Puebla,
Apartado Postal J-48, Puebla 72570, Mexico}

\author{Victor A. Gopar}
\affiliation{Depto de F\'isica Te\'orica, Facultad de Ciencias, and Instituto de Biocomputaci\'on y F\'isica de Sistemas Complejos (BIFI), Universidad de Zaragoza, Pedro Cerbuna 12, E-50009 Zaragoza, Spain}

\author{Imre Varga}
\affiliation{Elm\'eleti Fizika Tansz\'ek, Fizikai
Int\'ezet, Budapesti M\H uszaki \'es Gazdas\'agtudom\'anyi Egyetem,
H-1521 Budapest, Hungary}
\affiliation{Fachbereich Physik und Wissenschaftliches Zentrum f\"ur
Materialwissenschaften, Philipps Universit\"at Marburg, D-35032 Marburg,
Germany}

\date{\today}

\begin{abstract}
We study numerically scattering and transport statistical properties
of the one-dimensional Anderson model at the metal-insulator transition
described by the Power-law Banded Random Matrix (PBRM) model at
criticality. Within a scattering approach to electronic transport,
we concentrate on the case of a small number of single-channel
attached leads. We observe a smooth crossover from localized to
delocalized behavior in the average scattering matrix elements, the
conductance probability distribution, the variance of the
conductance, and the shot noise power by varying $b$ (the effective
bandwidth of the PBRM model) from small ($b\ll 1$) to large ($b>1$)
values. We contrast our results with analytic random matrix theory
predictions which are expected to be recovered in the limit $b\to \infty$.
We also compare our results for the PBRM model with those for the
three-dimensional (3D) Anderson model at criticality, finding that
the PBRM model with $b \in [0.2,0.4]$ reproduces well the scattering
and transport properties of the 3D Anderson model.
\end{abstract}


\maketitle


\section{Introduction}
The study of systems at the Anderson metal-insulator transition
(MIT) has been a subject of intensive research activity for several decades.~\cite{A58,AMPJ95,AKL91,EM08}
In particular, much interest has been focused on the scattering
properties of critical systems by analysing the probability
distribution functions of the resonance widths $\Gamma$ and Wigner
delay times
$\tau_{\tbox W}$,~\cite{MK05,OF05,MFME06,OKG03,F03,TC99,KW02,KW02b,WMK06,MI06}
as well as the transmission or dimensionless conductance $T$.~\cite{S90,M94,M99,SO97,SOK00,SMO00,WLS98,RS01,RMS001,TM02,SM05,JMZ99}
The distribution functions of $\Gamma$ and $\tau_{\tbox W}$
have been shown to be related to the properties of the
corresponding closed systems, i.e., the fractality of the eigenstates
and the critical features of the MIT. On the other hand, at the MIT,
the distribution of conductances $w(T)$ has been found to be {\it
universal}, i.e., size independent, but dependent on the adopted model,
dimensionality, symmetry, and even boundary conditions of the system.
$w(T)$ has been studied for systems in two and more dimensions with a
large number of attached single-channel
leads.~\cite{S90,M94,M99,SO97,SOK00,SMO00,WLS98,RS01,RMS001,TM02,SM05,JMZ99}
In fact, concerning the conductance of one-dimensional
(1D) systems and its statistical distribution,
the regime of small number of leads has been left almost unexplored.~\cite{JMZ99,SZ02,MG09a,MG09b}

In the present work we study numerically several statistical properties
of the scattering matrix and the electronic transport across disordered
systems in one and three dimensions described by the Power-law Banded
Random Matrix (PBRM) model at criticality and the three-dimensional (3D)
Anderson model at
the MIT, respectively. We stress that we concentrate on the case of a
small number of attached leads each of them supporting one open channel.

The organization of this paper is as follows.
In the next subsection we define the PBRM model, the 3D Anderson model,
as well as the scattering setup. We also define the scattering quantities
under investigation and provide the corresponding analytical predictions
from random scattering-matrix theory (RMT) for systems with
time-reversal symmetry. These analytical results will be used as a
reference along the paper. In Section II we analyse the average
scattering matrix elements, the conductance probability distribution,
the variance of the conductance, and the shot noise power for the PBRM
model as a function of its effective bandwidth $b$. In Section III we
compare the results of the PBRM model at criticality with the scattering
and transport properties of the 3D Anderson model.
Finally, Section IV is left for conclusions.

\subsection{The PBRM and the 3D Anderson models}

As we have mentioned above, in the present study we adopt two models,
namely, the PBRM model at criticality and the 3D Anderson model at the
MIT.

The PBRM model~\cite{EM08,MFDQS96} describes 1D samples of length $L$
with random long-range hopping. This model is represented by $N\times N$
($N=L$) real symmetric matrices whose elements are statistically
independent random variables drawn from a normal distribution with zero
mean, $\langle H_{ij} \rangle=0$, and a variance decaying as a power law
$\langle  |H_{ij}|^2 \rangle \sim (b/|i-j|)^{2\alpha}$, where $b$ and
$\alpha$ are parameters. There are two prescriptions for the variance
of the PBRM model: the so-called non-periodic,
\begin{equation}
   \langle  |H_{ij}|^2 \rangle =
   \frac{1}{2} \frac{1 + \delta_{ij}}{1+\left( |i-j|/b \right)^{2\alpha}} \ ,
\label{PBRMnp}
\end{equation}
where the 1D sample is in a {\it line} geometry; and the periodic,
\begin{equation}
   \langle  |H_{ij}|^2 \rangle =
   \frac{1}{2} \frac{1 + \delta_{ij}}{1+\left[
   \sin\left( \pi|i-j|/L \right)/(\pi b/L) \right]^{2\alpha}} \ ,
\label{PBRMp}
\end{equation}
where the sample is in a {\it ring} geometry.
Field-theoretical considerations~\cite{EM08,MFDQS96,KT00} and detailed
numerical investigations~\cite{EM08,EM00b,V03} have verified that the PBRM
model undergoes a transition at $\alpha=1$ from localized states for
$\alpha >1$ to delocalized states for $\alpha < 1$. This transition
shows all the key features of the Anderson MIT, including
multifractality of eigenfunctions and non-trivial spectral statistics
at the critical point. Thus the PBRM model possesses a line of
critical points $b\in (0,\infty)$. We set $\alpha=1$ in our study,
i.e., we work with the PBRM model at criticality.

The 3D Anderson model with diagonal disorder is described by the
tight-binding Hamiltonian (TBH)
\begin{equation}
\label {tbh}
H=\sum_{\bf n} |{\bf n}\rangle W_{\bf n}\langle {\bf n}| + \sum_{{\bf (n,m)}}
|{\bf n}\rangle \langle {\bf m}| \ ,
\end{equation}
where ${\bf n}\equiv (n_x,n_y,n_z)$ labels all the $N=L^3$ sites of
a cubic lattice with linear size $L$, while the second sum is taken
over all nearest-neighbour pairs ${\bf (n,m)}$ on the lattice. The
on-site potentials $W_{\bf n}$ for $1\leq n_x,n_y,n_z\leq L$ are
independent random variables. When $W_{\bf n}$ are Gaussian
distributed, with zero mean and variance $W^2/12$, the MIT at energy
$E\simeq 0$ occurs for $W=W_c\simeq 21.3$. See
Refs.~[\onlinecite{M99,RMS01,MW99}]. Then, for $W<W_c$ ($W>W_c$) the
system is in the metallic (insulating) regime. We set $W=W_c$ in our
study.

We open the isolated samples, defined above by the PBRM model
and the 3D Anderson model, by attaching $2M$ semi-infinite
single channel leads. Each lead is described by the 1D
semi-infinite TBH
\begin{equation}
\label{leads}
H_{\tbox{lead}}=\sum^{-\infty}_{n=1} (| n \rangle \langle n+1| + |n+1 \rangle \langle n|) \ .
\end{equation}

Using standard methods~\cite{MW69} one can write the scattering
matrix ($S$-matrix) in the form~\cite{SOKG00,KW02}
\begin{equation}
\label{smatrix}
S(E) =
\left(
\begin{array}{cc}
r & t'   \\
t & r'
\end{array}
\right)
={\bf 1}-2i \sin (k)\, {\cal W}^{\,T} (E-{\cal H}_{\rm eff})^{-1} {\cal W} \ ,
\end{equation}
where ${\bf 1}$ is the $2M\times 2M$ unit matrix, $k=\arccos(E/2)$ is
the wave vector supported in the leads, and ${\cal H}_{\rm eff}$ is
an effective non-hermitian Hamiltonian given by
\begin{equation}
\label{Heff}
{\mathcal{H}}_{\rm eff}=H- e^{ik} {\cal W}{\cal W}^{\,T} \ .
\end{equation}
Here, ${\cal W}$ is an $N\times 2M$ matrix that specifies the positions
of the attached leads to the sample. Its elements are equal to zero or
$\epsilon$, where $\epsilon$ is the coupling strength. Moreover, assuming
that the wave vector $k$ do not change significantly in the centre of the
band, we set $E=0$ and neglect the energy dependence of
${\mathcal{H}}_{\rm eff}$ and $S$.

\subsection{RMT predictions for the Circular Orthogonal Ensemble}

Notice from Eqs.~(\ref{PBRMnp}-\ref{PBRMp}) that in the limit
$b\to \infty$ the PBRM model reproduces the Gaussian Orthogonal
Ensemble. Therefore, in that limit we expect the statistics of
the scattering matrix, Eq.~(\ref{smatrix}), to be determined by the
Circular Orthogonal Ensemble (COE) which is the appropriate ensemble
for systems with time reversal symmetry.
Thus, below, we provide the statistical results for the
average $S$-matrix and the transport quantities to be analysed
in the following sections, assuming the orthogonal symmetry.
In all cases, we also assume the absence of direct processes, i.e.,
$\langle S \rangle =0$.

We start with the average of the $S$-matrix elements.
It is known that
\begin{equation}
\label{coeS}
\bra |S_{aa'}|^2 \ket_{\tbox{COE}} = \frac{1+\delta_{aa'}}{2M+1} \ ,
\end{equation}
where $\bra \cdot \ket$ means ensemble average over the COE.

Within a scattering approach to the electronic transport, once the
scattering matrix is known one can compute the dimensionless
conductance $T={\mbox{Tr}}(tt^\dagger)$ and its distribution $w(T)$.
For $M=1$, i.e., considering two single-channels leads attached to
the sample, $w(T)$ is given by
\begin{equation}
\label{GOEN1}
w(T)_{\tbox{COE}} = \frac{1}{2\sqrt{T}} \ ,
\end{equation}
while for $M=2$,
\begin{equation}
w(T)_{\tbox{COE}} = \left\{
\label{GOEN2}
\begin{array}{ll}
\frac{3}{2} T \ , & 0<T<1  \\
\frac{3}{2} \left( T-2\sqrt{T-1}\right) \ , & 1<T<2 \\
\end{array} \right. \ . \\
\end{equation}
For arbitrary $M$, the predictions for the average value of $T$
and its variance are
\begin{equation}
\label{GOEavT}
\bra T \ket_{\tbox{COE}} = \frac{M}{2}-\frac{M}{2(2M+1)} \
\end{equation}
and
\begin{equation}
\label{GOEvarT}
\mbox{var}(T)_{\tbox{COE}} = \frac{M(M+1)^2}{(2M+1)^2(2M+3)} \ ,
\end{equation}
respectively. For the derivation of the expressions above see for
example Ref.~[\onlinecite{MK04}].
Another transport quantity of interest is the shot noise
power $P = \bra {\mbox{Tr}}(tt^\dagger - tt^\dagger tt^\dagger) \ket$,
which as a function of $M$ reads~\cite{evgeny,HMBH06,savin}
\begin{equation}
\label{GOEP}
P_{\tbox{COE}} = \frac{M(M+1)^2}{2(2M+1)(2M+3)} \ .
\end{equation}

In the following sections we focus on $\bra |S_{aa'}|^2 \ket$,
$\bra T \ket$, $\mbox{var}(T)$, and $P$ for the PBRM model and the 3D
Anderson model, both at the MIT. In all cases we set the coupling strength
$\epsilon$ such that $\bra S \ket \approx 0$ in order to compare our
results, in the proper limits, with the RMT predictions introduced above.

\section{PBRM model}

\begin{figure}[t]
\centerline{\includegraphics[width=5cm]{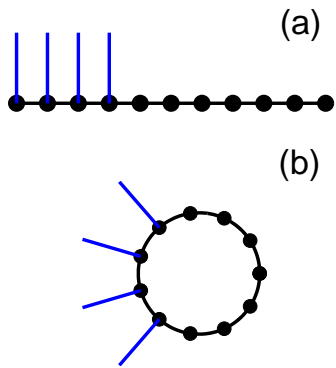}}
\caption{(Color online) Scattering setup.
$2M$ leads, shown as blue (gray) lines,
are attached to $2M$ sites (black dots)
of a 1D sample described by the (a) non-periodic and (b)
periodic versions of the PBRM model. The case $M=2$ with
$L=11$ is shown as example.}
\label{setup}
\end{figure}

We attach the $2M$ leads to the first $2M$ sites of the 1D sample
described by the PBRM model.
That is, in the non-periodic version of the PBRM model,
Eq.~(\ref{PBRMnp}), we attach the leads at the {\it boundary} of
the system. See Fig.~\ref{setup}(a). While in the periodic version,
Eq.~(\ref{PBRMp}), we attach them to the {\it bulk}. See Fig.~\ref{setup}(b).
In the latter case, finite size effects are considerably reduced.
However, the quantities we analyse below are $L$-independent once
$L$ is much larger than the number of attached leads for both
versions of the PBRM model.

We point out that our setup is significantly different from the
one used in Refs.~[\onlinecite{MG09a,MG09b}], where the conductance
has also been studied using the PBRM model.
There, for example, the leads in the $M=1$ case are attached to
sites which are separated a distance of $L/2$ and $L$ in the
periodic and non-periodic versions of the PBRM model, respectively.
In such situation the scattering quantities are strongly
$L$-dependent.

\begin{figure}[h]
\centerline{\includegraphics[width=8cm]{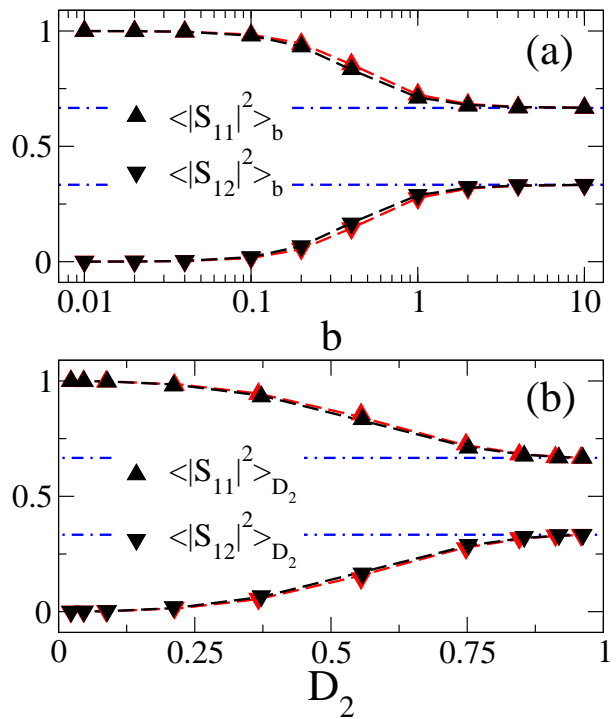}}
\caption{(Color online) Black [red (gray)] symbols: Average $S$-matrix
elements $\bra |S_{11}|^2 \ket$ and $\bra |S_{12}|^2 \ket$
for the periodic [non-periodic] PBRM model at criticality
as a function of (a) $b$ and (b) $D_2$ for $M=1$.
The blue (gray) dot-dashed lines correspond to 2/3 and 1/3;
the RMT prediction for $\bra |S_{11}|^2 \ket$ and
$\bra |S_{12}|^2 \ket$, respectively, given by Eq.~(\ref{coeS}).
The black [red (gray)] dashed lines in (a) are Eqs.~(\ref{S11b})
and (\ref{S12b}) with $\delta \approx 2.5$ [$\delta \approx 2.2$].
The black [red (gray)] dashed lines in (b) are Eqs.~(\ref{S11D}) and
(\ref{S12D}) with $\delta \approx 2.5$ [$\delta \approx 2.2$].
Error bars are not shown since they are much smaller
than symbol size.}
\label{Fig1}
\end{figure}

\subsection{Average scattering matrix elements}

First we consider the case $M=1$, where the $S$-matrix is a
$2\times 2$ matrix. In Fig.~\ref{Fig1}(a) we plot the ensemble
average of the elements $|S_{11}|^2$ and $|S_{12}|^2$ as a
function of the bandwidth parameter $b$, $\bra |S_{11}|^2 \ket_b$
and $\bra |S_{12}|^2 \ket_b$, for the periodic (black symbols)
and the non-periodic (red symbols) PBRM model.
We concentrate on these two matrix elements since the
other two elements give no additional information:
$\bra |S_{22}|^2 \ket_b = \bra |S_{11}|^2 \ket_b$ and
$\bra |S_{21}|^2 \ket_b = \bra |S_{12}|^2 \ket_b$.
Notice a strong $b$-dependence of the average $S$-matrix
elements driving them from a localized-like regime
[$\bra |S_{11}|^2 \ket \approx 1$ and
$\bra |S_{12}|^2 \ket \approx 0$; i.e. the average conductance
is close to zero] for $b\ll 1$, to a delocalized-like or
ballistic-like regime [$\bra |S_{11}|^2 \ket \approx 2/3$ and
$\bra |S_{12}|^2 \ket \approx 1/3$; i.e., RMT results are already
recovered] for $b \ge 4$.

Moreover, we have found that $\bra |S_{11}|^2 \ket$ and
$\bra |S_{12}|^2 \ket$, as a function of the bandwidth $b$,
are well described by
\begin{eqnarray}
\label{S11b}
\bra |S_{11}|^2 \ket_b & = &
1 - \bra |S_{12}|^2 \ket_b \ , \\
\label{S12b}
\bra |S_{12}|^2\ket_b & = &
\frac{1}{3} \left[ \frac{1}{1+(\delta b)^{-2}}
\right] \ ,
\end{eqnarray}
where $\delta$ is a fitting parameter. Eq.~(\ref{S11b}) is a
consequence of the unitarity of the scattering matrix,
$SS^\dagger =\mathbf 1$, while the factor 1/3 in Eq.~(\ref{S12b})
comes from Eq.~(\ref{coeS}) with $M=1$.
In Fig.~\ref{Fig1}(a) we plot Eqs.~(\ref{S11b}) and (\ref{S12b})
(dashed lines) and compare them with the corresponding results
from the PBRM model (symbols) in the periodic and non-periodic
setups. In the same figure Eq.~(\ref{coeS}) is also plotted
(dot-dashed lines).

On the other hand, it is well known that in systems at the disorder
driven MIT both the energy spectra and the eigenstates exhibit
multifractal characteristics.~\cite{EM08} The PBRM model is
characterized by the effective bandwidth $b$ that drives the system
from strong ($b\to 0$) to weak ($b\to \infty$) multifractality.
Multifractality can be quantified by the generalized dimensions
$D_q$ which describe the fluctuations of the eigenfunctions.
The multifractal dimensions $D_q$ of the $\sigma$-th eigenfunction
$\Psi^\sigma$ (given as a linear combination of the basis states in
a system with linear size $L$,
$\Psi^\sigma = \sum_{l=1}^{L} C_l^\sigma \phi_l$)
are defined through the so-called inverse participation numbers,
${\cal I}_q = \sum_{l=1}^L \left| C_l^\sigma \right|^{2q}$, by the
scaling~\cite{EM08,EM00a,V03}
\begin{equation}
\label{Dq}
    ({\cal I}_q)^{\tbox{typ}} \propto L^{-(q-1)D_q} \ ,
\end{equation}
where $({\cal I}_q)^{\tbox{typ}} = \exp \bra \ln {\cal I}_q \ket$ is
the typical value of ${\cal I}_q$.
However, among all dimensions, the correlation dimension $D_2$ plays
a prominent role.~\cite{HP83} As $b$ transits from zero to infinity,
$D_2$ takes values from zero to one.

Using numerical diagonalization of
matrices with sizes $L=128$, 256, 512, and 1024, we extracted $D_2$
by the use of Eq.~(\ref{Dq}). For each system size $L$, we used a number
of disorder realizations in order to have at least $10^5$ data for
its analysis. For each disorder realization we used $25\%$
of the states at the centre of the spectral band (where the density of
states is almost constant) in order to avoid boundary effects.
We found good agreement between the numerically obtained
$D_2$ and the analytical estimation~\cite{EM08,EM00a}
\begin{equation}
\label{D2}
   D_2 =\left\{
   \begin{array}{ll}
                    2b \ ,             & \quad b\ll 1 \nonumber\\
                    1-(\pi b)^{-1} \ , & \quad b\gg 1
   \end{array}
\right. \ .
\end{equation}
Moreover, in Ref.~[\onlinecite{MKC05}] the following phenomenological
analytical expression for $D_2$ as a function of $b$ was proposed:
\begin{equation}
\label{D2ofb}
    D_2(b) = \frac{1}{1+(\gamma b)^{-1}} \ ,
\end{equation}
where $\gamma$ is a fitting parameter. See also Ref.~[\onlinecite{MI06}].
Eq.~(\ref{D2ofb}) describes well our numerical results for $D_2$ with
$\gamma=2.94$ and $\gamma=2.88$ for the periodic and non-periodic
versions of the PBRM model, respectively.

\begin{figure}[t]
\centerline{\includegraphics[width=8cm]{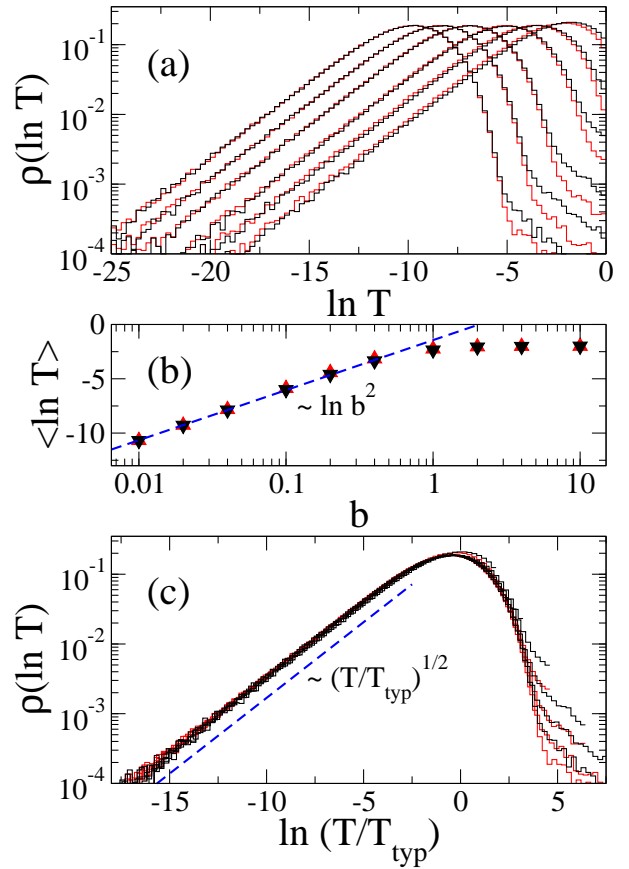}}
\caption{(Color online) (a) Black [red (gray)] curves:
Probability distribution $\rho(\ln T)$ for the periodic
[non-periodic] PBRM model at criticality for several values
of $b < 1$ ($b=0.01$, 0.02, 0.04, 0.1, 0.2, and 0.4 from left
to right) in the case $M=1$.
(b) Black [red (gray)] symbols: $\bra \ln T \ket$ as a function of
$b$ for the periodic [non-periodic] PBRM model. The blue (gray)
dashed line is the best fit of the data to the logarithmic
function $A + \ln b^2$ with $A\approx-1.44$.
(c) $\rho(\ln T)$ for $b < 1$ scaled to
$T_{\tbox{typ}}=\exp\langle \ln T \rangle \sim b^2$.
The blue (gray) dashed line proportional to $(T/T_{\tbox{typ}})^{1/2}$
is plotted to guide the eye.}
\label{Fig2}
\end{figure}
\begin{figure}[ht]
\centerline{\includegraphics[width=8cm]{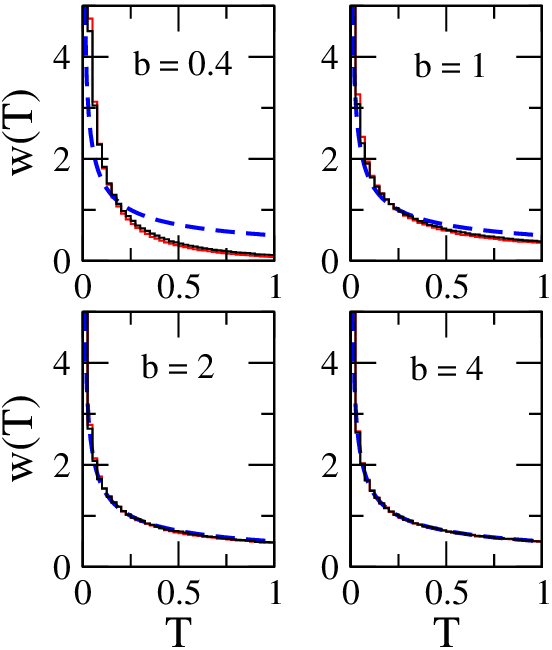}}
\caption{(Color online) Black [red (gray)] curves: Conductance
probability distribution $w(T)$ for the periodic
[non-periodic] PBRM model at criticality for some
{\it large} values of $b$ in the case $M=1$.
Blue (gray) dashed lines are $w(T)_{\tbox{COE}}$; the RMT
prediction for $w(T)$ given by Eq.~(\ref{GOEN1}).}
\label{Fig3}
\end{figure}
\begin{figure}[htb]
\centerline{\includegraphics[width=8cm]{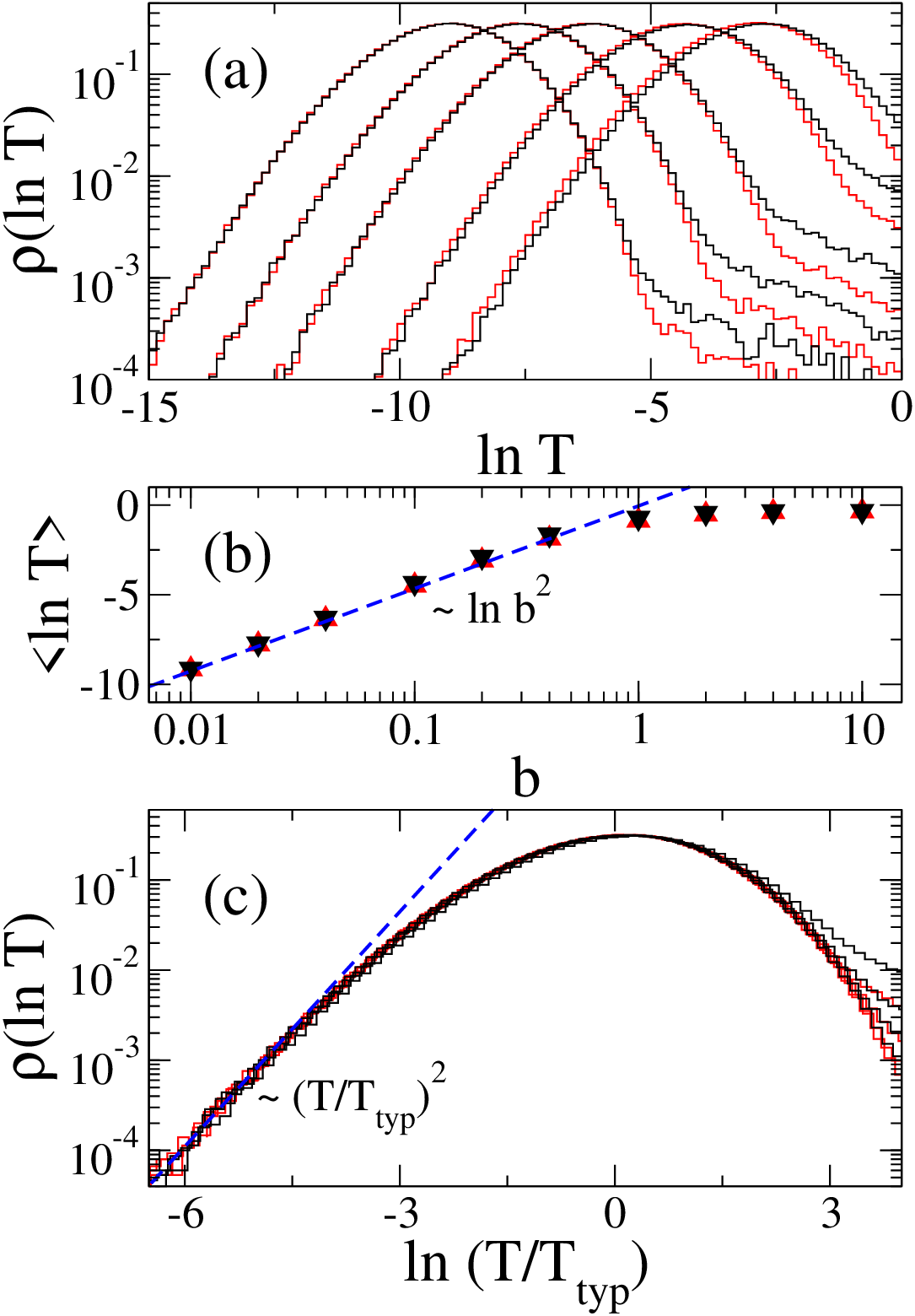}}
\caption{(Color online) (a) Black [red (gray)] curves:
Probability distribution $\rho(\ln T)$ for the periodic
[non-periodic] PBRM model at criticality for several values
of $b < 1$ ($b=0.01$, 0.02, 0.04, 0.1, and 0.2 from left
to right) in the case $M=2$.
(b) Black [red (gray)] symbols: $\bra \ln T \ket$ as a function of
$b$ for the periodic [non-periodic] PBRM model. The blue (gray)
dashed line is the best fit of the data to the logarithmic
function $A + \ln b^2$ with $A\approx -0.057$.
(c) $\rho(\ln T)$ for $b < 1$ scaled to
$T_{\tbox{typ}}=\exp\langle \ln T \rangle \sim b^2$.
The blue (gray) dashed line proportional to $(T/T_{\tbox{typ}})^{2}$
is plotted to guide the eye.}
\label{Fig5}
\end{figure}
\begin{figure}[htb]
\centerline{\includegraphics[width=8cm]{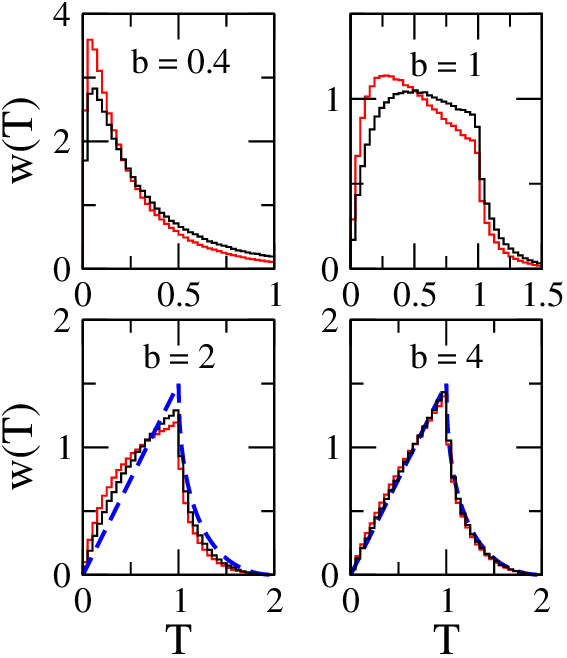}}
\caption{(Color online) Black [red (gray)] curves: Conductance
probability distribution $w(T)$ for the periodic
[non-periodic] PBRM model at criticality for some
{\it large} values of $b$ in the case $M=2$.
Blue (gray) dashed lines are $w(T)_{\tbox{COE}}$; the RMT
prediction for $w(T)$ given by Eq.~(\ref{GOEN2}).}
\label{Fig6}
\end{figure}
\begin{figure}[htb]
\centerline{\includegraphics[width=8cm]{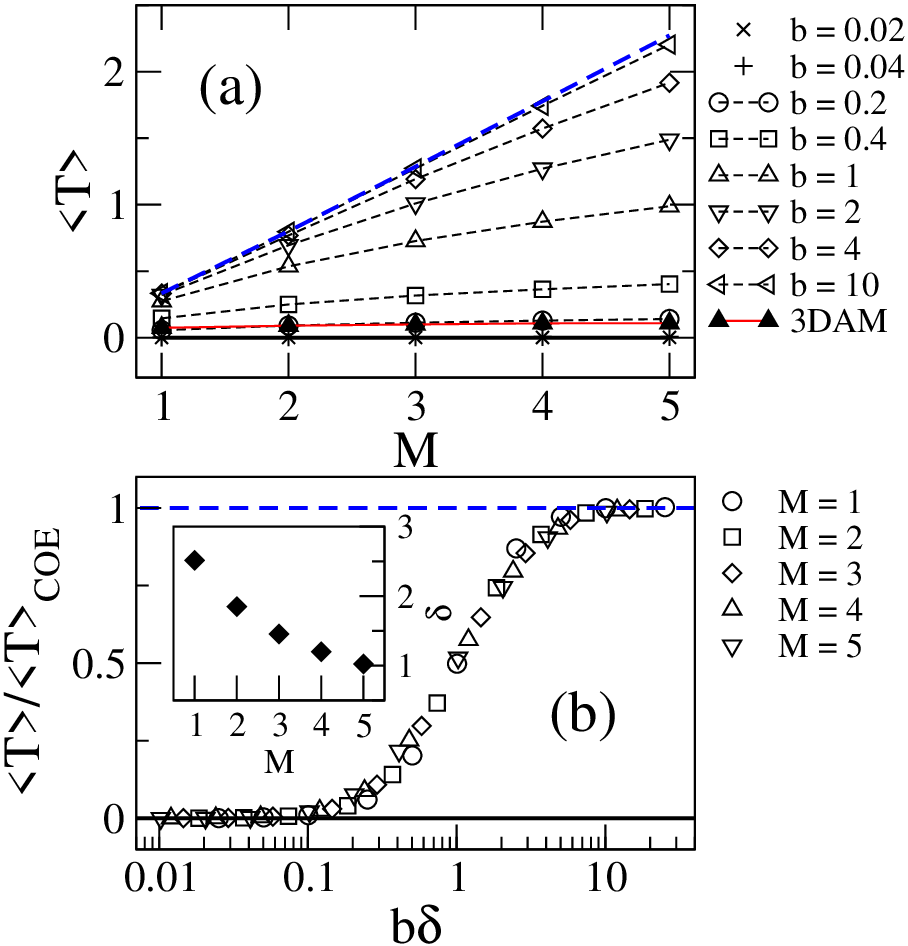}}
\caption{(Color online) (a) Average conductance $\bra T \ket$
as a function of $M$ for the non-periodic PBRM model at
criticality for several values of $b$. $\bra T \ket$ for the
3D Anderson model at criticality (3DAM) is also shown.
(b) $\bra T \ket/\bra T \ket_{\tbox{COE}}$ as a function
of $b\delta$ for the periodic PBRM model at criticality
for $M\in[1,5]$. Insert: $\delta$ versus $M$.
$\delta$ is obtained from the fitting of Eq.~(\ref{Xb})
to the $\bra T \ket$ vs. $b$ data.
Thick full lines correspond to $\bra T \ket = 0$.
Blue (gray) dashed lines are (a) the RMT prediction for
$\bra T \ket$ given by Eq.~(\ref{GOEavT}); and (b) one.}
\label{Fig7}
\end{figure}
\begin{figure}[htb]
\centerline{\includegraphics[width=8cm]{Fig8.eps}}
\caption{(Color online) (a) Variance of $T$
as a function of $M$ for the non-periodic PBRM model at
criticality for several values of $b$. $\mbox{var}(T)$ for the
3D Anderson model at criticality (3DAM) is also shown.
(b) $\mbox{var}(T)/\mbox{var}(T)_{\tbox{COE}}$ as a function
of $b\delta$ for the periodic PBRM model at criticality
for $M\in[1,5]$. Insert: $\delta$ versus $M$.
$\delta$ is obtained from the fitting of Eq.~(\ref{Xb})
to the $\mbox{var}(T)$ vs. $b$ data.
Thick full lines correspond to $\mbox{var}(T) = 0$.
Blue (gray) dashed lines are (a) the RMT prediction for
$\mbox{var}(T)$ given by Eq.~(\ref{GOEvarT}); and (b) one.}
\label{Fig8}
\end{figure}
\begin{figure}[htb]
\centerline{\includegraphics[width=8cm]{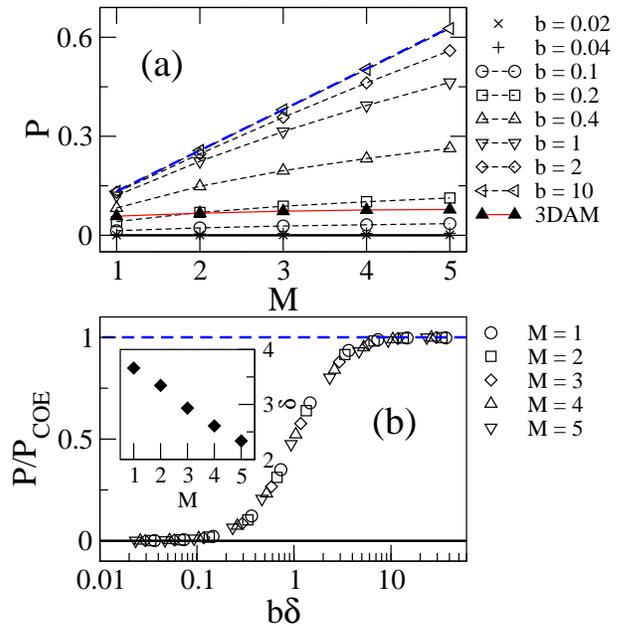}}
\caption{(Color online) (a) Shot noise power $P$
as a function of $M$ for the non-periodic PBRM model at
criticality for several values of $b$. $P$ for the
3D Anderson model at criticality (3DAM) is also shown.
(b) $P/P_{\tbox{COE}}$ as a function
of $b\delta$ for the periodic PBRM model at criticality
for $M\in[1,5]$. Insert: $\delta$ versus $M$.
$\delta$ is obtained from the fitting of Eq.~(\ref{Xb})
to the $P$ vs. $b$ data.
Thick full lines correspond to $P = 0$.
Blue (gray) dashed lines are (a) the RMT prediction for
$P$ given by Eq.~(\ref{GOEP}); and (b) one.}
\label{Fig9}
\end{figure}

Then, by substituting Eq.~(\ref{D2ofb}) into Eqs.~(\ref{S11b})
and (\ref{S12b}) we obtain the following expressions for the
averages of $|S_{11}|^2$ and $|S_{12}|^2$ as a function of $D_2$:
\begin{eqnarray}
\label{S11D}
\bra |S_{11}|^2 \ket_{D_2} & = &
1 - \bra |S_{12}|^2 \ket_{D_2} \ , \\
\label{S12D}
\bra |S_{12}|^2 \ket_{D_2} & = &
\frac{1}{3}\frac{1}
{1+(\gamma/\delta)^2\left(D_2^{-1}-1\right)^2} \ .
\end{eqnarray}
In Fig.~\ref{Fig1}(b) we compare Eqs.~(\ref{S11D}) and
(\ref{S12D}) (dashed lines) with the ensemble-average
numerical results at different values of $D_2$.
The agreement is excellent.
In addition, it is interesting to note that
\begin{equation}
\label{S12}
   \bra |S_{12}|^2 \ket_{D_2} \propto \left\{
   \begin{array}{ll}
                    D_2^2 \ ,     & \quad D_2\to 0 \nonumber\\
                    1-(\gamma/\delta)^2\left(D_2^{-1}-1\right)^2 \ , & \quad D_2\to 1
   \end{array}
\right. \ ,
\end{equation}
which might be relevant for systems at the MIT where $D_2$ can be
tuned.

Finally we want to remark that concerning $\bra |S_{aa'}|^2 \ket$
for the PBRM model, the RMT limit, expected for $b\to \infty$,
is already recovered for $b\ge 4$. See Fig.~\ref{Fig1}(a).

\begin{table}[b]
\begin{ruledtabular}
\begin{tabular}{|l|l|l|l|}
$L$ &  $b = 0.1$ & $b = 1$ & $b = 10$ \\ \hline
50  & $-5.887\pm 0.002$ & $-2.249\pm 0.002$ & $-2.001\pm 0.002$\\
100 & $-5.886\pm 0.002$ & $-2.247\pm 0.002$ & $-1.999\pm 0.002$\\
200 & $-5.875\pm 0.002$ & $-2.245\pm 0.002$ & $-1.999\pm 0.002$\\
400 & $-5.875\pm 0.002$ & $-2.244\pm 0.002$ & $-2.002\pm 0.002$\\
800 & $-5.869\pm 0.002$ & $-2.228\pm 0.002$ & $-1.989\pm 0.002$
\end{tabular}
\end{ruledtabular}
\caption{$\bra \ln T \ket$ for the periodic PBRM model at
criticality for some values of $b$ and $L$ in the case $M=1$.}
\label{TablePBRM}
\end{table}

\subsection{Conductance probability distribution}

Now we turn to the conductance statistics.
For $b\ll 1$ the conductance distribution $w(T)$ is highly
concentrated close to $T=0$. So it is more convenient to
analyse the distribution of the transmission logarithm,
$\rho(\ln T)$, instead. Then, in Fig.~\ref{Fig2}(a) we show
$\rho(\ln T)$ for several values of $b<1$ in the case $M=1$,
for the periodic and non-periodic versions of the PBRM model.
Notice that the distribution functions $\rho(\ln T)$ do not
change their shape  by increasing $b$, mainly for $b\ll 1$,
thus being scale invariant. In fact, $\bra \ln T \ket$ for
$b<1$ clearly displays a linear behavior when plotted as
a function of $\ln b$ as shown in Fig.~\ref{Fig2}(b). Then,
all distributions functions $\rho(\ln T)$ fall one on top of the
other when shifting them along the $x$-axis by the typical
value of $T$,
\begin{equation}
T_{\tbox{typ}} = \exp\langle \ln T \rangle \ ,
\end{equation}
as shown in Fig.~\ref{Fig2}(c).

Note that for $T<T_{\tbox{typ}}$, $\rho(\ln T)$ is proportional
to $(T/T_{\tbox{typ}})^{1/2}$, see Figs.~\ref{Fig2}(a) and
\ref{Fig2}(c). Moreover, we found that such behavior extends
from small to large values of $b$, $b\gg 1$.
Also, notice that the behavior $\rho(\ln T)\propto T^{1/2}$ for small $T$
coincides with the RMT prediction $w(T)_{\tbox{COE}}\propto T^{-1/2}$,
see Eq.~(\ref{GOEN1}), since the change of variable $T \to \ln T$
leads to $\rho(\ln T)=T w_{\tbox{COE}}(T)$.

In Fig.~\ref{Fig3} we show $w(T)$ for large $b$ ($b\ge 0.4$).
In the limit $b\to \infty$, $w(T)$ is expected
to approach the RMT prediction of Eq.~(\ref{GOEN1}).
However, once $b\ge 4$, $w(T)$ is already well described by
$w(T)_{\tbox{COE}}$.

We want to stress that our results do not depend on the system
size $L$, as can be seen in Table~\ref{TablePBRM} where we report
$\bra \ln T \ket$ for the periodic PBRM model at criticality for
some values of $b$ and $L$ in the case $M=1$.
In fact, $\bra \ln T \ket$ does not depend on $L$ out of
criticality ($\alpha>1$ and $\alpha<1$) either.
This makes an important difference with respect to the
setup where the leads are attached to opposite sides of the
sample where $\bra \ln T \ket$ increases (decreases) as a function 
of $L$ for $\alpha < 1$ ($\alpha > 1$ and $\alpha = 1$).

Now, in Figs.~\ref{Fig5} and \ref{Fig6} we explore $w(T)$
in the case $M=2$. As well as in the case $M=1$, studied above,
here: (i) for small $b$, $\rho(\ln T)$ is scale invariant with
$T_{\tbox{typ}}$ as scaling factor, see Fig.~\ref{Fig5};
(ii) for $b\ge 4$, $w(T)$ is well described by Eq.~(\ref{GOEN2}),
the corresponding RMT prediction, see Fig.~\ref{Fig6}.
However, although $\rho(\ln T)$ for small $b$ and $w(T)$ for
large $b$ are practically the same for the periodic and
non-periodic versions of the PBRM model, they show differences
for intermediate values, $0.1<b<4$, as can be seen in
Fig.~\ref{Fig5}(a) and the upper panels of Fig.~\ref{Fig6}.

We point out that for small $T$, $T\ll T_{\tbox{typ}}$,
$\rho(\ln T)$ is proportional to $(T/T_{\tbox{typ}})^{2}$, as
shown in Fig.~\ref{Fig5}(c). This behavior is
{\it universal} for the PBRM model with $M=2$; i.e. it is
$b$-independent and valid for the periodic and non-periodic
versions of the model. Again, as for the $M=1$ case, here for
$M=2$, the dependence $\rho(\ln T)\propto T^2$ in the limit
$b\to \infty$ is consistent with the RMT prediction
$w(T)_{\tbox{COE}} \propto T$ for $T<1$; see Eq.~(\ref{GOEN2}).

From the results above and since~\cite{MK04}
\begin{equation}
\label{wMRMT}
w(T)_{\tbox{COE}} \propto T^{M^2/2-1} \ ,
\end{equation}
in the region $0<T<1$, we conclude that for the PBRM model
\begin{equation}
\label{wM}
\rho(\ln T) \propto T^{M^2/2} \ ,
\end{equation}
for $T\ll T_{\tbox{typ}}$.
We argue that the behavior dictated by Eq.~(\ref{wM}) can
be interpreted in the following way: Since small values of
conductance $T\ll 1$ mean very strong reflection, the
scattering process is almost direct, i.e. an incident
electron is scattered out the system mainly by the
interaction with the sites at which the leads are attached.
Then the electron does not explore the complete sample,
not even part of it, and as a consequence it does not realize
that the sample is at criticality.
That is, for $T\ll 1$, the incident electron does not distinguish
between a critical system and a random system represented by
a full random matrix if the number of attached leads $2M$ is
not much larger than $b$; i.e., when the leads are attached
to sites interconnected by Hamiltonian matrix elements located
within the bandwidth of the matrix where all elements have almost
the same variance: $\bra |H_{ij}|^2 \ket \approx 1/2$.
So, we can use Eq.~(\ref{wMRMT}) and as a consequence
Eq.~(\ref{wM}) too.
In fact, we have verified that Eq.~(\ref{wM}) is valid up to
$M=2$ for $b\le 1$, while for $b=10$ it describes perfectly the
left tail of $\rho(\ln T)$ for all the values of $M$ used in this
work (up to $M=5$). Our result, given by Eq.~(\ref{wM}), is the
generalization of the result shown in Ref.~[\onlinecite{MG09a}],
where the behavior $\rho(\ln T) \propto T^{1/2}$ was reported for
the PBRM model in the localized ($\alpha>1$) and delocalized
regimes ($\alpha<1$) for $M=1$.

\subsection{Mean and variance of the conductance and the shot noise power}

\begin{figure}[t]
\centerline{\includegraphics[width=8cm]{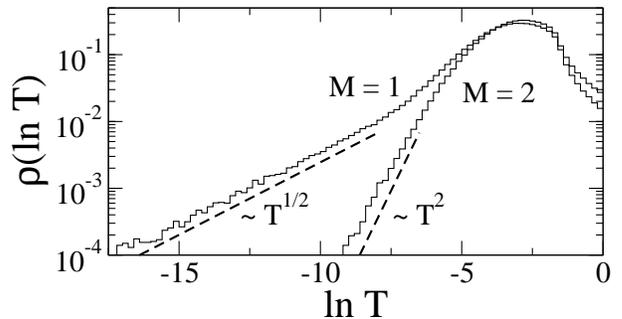}}
\caption{Probability distribution $\rho(\ln T)$ for the 3D
Anderson model at criticality in the cases
$M=1$ and $M=2$. Dashed lines with slopes $T^{1/2}$
and $T^{2}$ are plotted to guide the eye.}
\label{Fig10}
\end{figure}
\begin{figure}[t]
\centerline{\includegraphics[width=8cm]{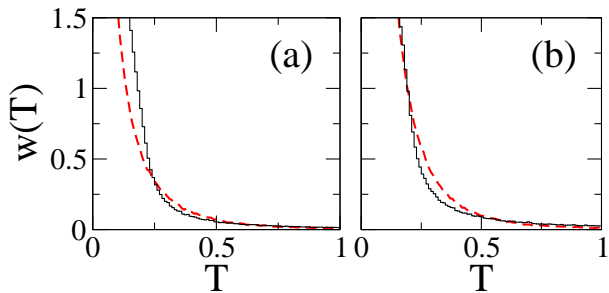}}
\caption{(Color online) Conductance probability distribution
$w(T)$ for the 3D Anderson model at criticality in the cases
(a) $M=1$ and (b) $M=2$. The red (gray) dashed line corresponds
to $w(T)$ for the non-periodic PBRM model with $b=0.2$.}
\label{Fig11}
\end{figure}

We now increase further the number of attached
leads.\footnote{All quantities reported in Figs.~\ref{Fig1}
to \ref{Fig6} were obtained for $L=50$ using $10^6$ ensemble
realizations. In Subsection IIC we used $L=200$ and $10^5$
ensemble realizations. We have verified that in both cases
our results are invariant by further increasing $L$. See
also Table~\ref{TablePBRM}.}
In Figs.~\ref{Fig7}(a), \ref{Fig8}(a), and \ref{Fig9}(a) we plot
the average conductance $\bra T \ket$, the variance of the
conductance $\mbox{var}(T)$, and the shot noise power $P$ for
the non-periodic version of the PBRM model for several values
of $b$ with $M\in [1,5]$.\footnote{
We recall that for $M=5$, ten single-channel leads are attached
to the sample.}
It is clear from these three plots that changing
$b$ from small ($b\ll 1$) to large ($b > 1$) values produces a
transition from localized- to delocalized-like behavior in the
scattering properties of the PBRM model. That is, (i) for
$b< 0.1$, $\bra T \ket \approx 0$, $\mbox{var}(T) \approx 0$,
and $P \approx 0$; and (ii) for $b\ge 10$, $\bra T \ket$,
$\mbox{var}(T)$, and $P$ are well described by the corresponding
RMT predictions given by Eqs.~(\ref{GOEavT}),
(\ref{GOEvarT}), and (\ref{GOEP}), respectively.
Similar plots are obtained (not shown here) for the periodic
PBRM model.

Moreover, we have observed that $\bra T \ket$, $\mbox{var}(T)$,
and $P$ behave (for all $M$) as $\bra |S_{12}|^2\ket_b$ does.
See Eq.~(\ref{S12b}). Thus, we can write
\begin{equation}
\label{Xb}
X(b) = X_{\tbox{COE}} \left[ \frac{1}{1+(\delta b)^{-2}}
\right] \ ,
\end{equation}
where $X$ represents $\bra T \ket$, $\mbox{var}(T)$, or $P$
and $\delta$ is a fitting parameter.
Then, in Figs.~\ref{Fig7}(b), \ref{Fig8}(b), and \ref{Fig9}(b)
we plot $\bra T \ket$, $\mbox{var}(T)$, and $P$ normalized to
their respective COE average values, now for the periodic
PBRM model as a function of $b\delta$ for $M\in [1,5]$.
Also, similar plots are obtained (not shown here) for the
non-periodic PBRM model.

\section{3D Anderson model}

Since the most prominent realization of systems that undergo a
MIT is the 3D Anderson model, it is of relevance to analyse its
scattering and transport properties taking as a reference the
results shown in the previous section for the PBRM model.

We attach the $2M$ leads to $2M$ sites at one of the edges of the
cubic lattice described by the 3D Anderson model. In this way we
make a {\it line contact} as we did in the case of the PBRM model.
So, we can compare the scattering properties of both models.
We use Gaussian distributed on-site potentials and
system sizes from $L=6$ to 10 (we have verified that our
results do not change by increasing $L$ further) with $10^6$ to
$10^4$ disorder realizations.

We start our analysis by looking at the average scattering matrix
elements for $M=1$. For the 3D Anderson model at criticality we found
$\bra |S_{11}|^2 \ket \approx 0.926$ and
$\bra |S_{12}|^2 \ket \approx 0.074$, see Table~\ref{Table1}, which
are close to those of the non-periodic PBRM model with $b=0.24$.

In Fig.~\ref{Fig10} we show conductance probability distributions
$\rho(\ln T)$ for the 3D Anderson model at criticality in the cases
$M=1$ and $M=2$. We found $\bra \ln T \ket \approx -3.47$ and $-3.06$
for $M=1$ and $M=2$, respectively. See Table~\ref{Table2}.
These values of $\bra \ln T \ket$
for the 3D Anderson model are close to those of the PBRM model with
$b=0.36$ and $b=0.22$, respectively.
Notice also that, as well as for the PBRM model, for the 3D Anderson
model $\rho(\ln T)$ in the region of  $T\ll T_{\tbox{typ}}$
is proportional to $(T/T_{\tbox{typ}})^{1/2}$ and
$(T/T_{\tbox{typ}})^{2}$ for $M=1$ and $M=2$, respectively
(dashed lines in Fig.~\ref{Fig10}).

Additionally, in Fig.~\ref{Fig11} we plot $w(T)$ for the 3D Anderson
model in the cases $M=1$ and $M=2$. In the same figure we have also
plotted (dashed lines) the conductance distributions from the PBRM
model with $b=0.2$. We can see that the conductance distributions of
both models are similar at this bandwidth value.

\begin{table}[t]
\begin{ruledtabular}
\begin{tabular}{|c|l|l|}
$L$ & $\bra |S_{11}|^2 \ket$ & $\bra |S_{12}|^2 \ket$ \\ \hline
6   & $0.9256\pm 0.0001$ & $0.0743\pm 0.0001$ \\
8   & $0.9254\pm 0.0003$ & $0.0745\pm 0.0003$ \\
10  & $0.926\pm 0.001$   & $0.073\pm 0.001$
\end{tabular}
\end{ruledtabular}
\caption{Average $S$-matrix elements $\bra |S_{11}|^2 \ket$ and
$\bra |S_{12}|^2 \ket$ for the 3D Anderson model at criticality
for some values of $L$ in the case $M=1$.}
\label{Table1}
\end{table}
\begin{table}[t]
\begin{ruledtabular}
\begin{tabular}{|c|l|l|}
$L$ & $\bra \ln T \ket (M=1)$ & $\bra \ln T \ket (M=2)$ \\ \hline
6   & $-3.471\pm 0.001$ & $-3.060\pm 0.001$ \\
8   & $-3.483\pm 0.005$ & $-3.058\pm 0.003$ \\
10  & $-3.450\pm 0.016$   & $-3.078\pm 0.012$
\end{tabular}
\end{ruledtabular}
\caption{$\bra \ln T \ket$ for the 3D Anderson model at
criticality for some values of $L$ in the cases $M=1$ and
$M=2$.}
\label{Table2}
\end{table}

We also compute $\bra T \ket$, $\mbox{var}(T)$, and $P$ for the 3D
Anderson model and plot them in Figs.~\ref{Fig7}(a), \ref{Fig8}(a),
and \ref{Fig9}(a), respectively (red curves labeled as 3DAM). We
observe that for the 3D Anderson model $\bra T \ket$,
$\mbox{var}(T)$, and $P$ behave as the corresponding
quantities for the PBRM model with $b$ close to 0.2.

We want to stress that our results for the 3D Anderson model
at criticality do not seem to depend on the on-site potential
distribution. Here we used Gaussian distributed potentials.
However, we observe practically the same results by the use
of box distributed on-site potentials (not shown here).

\section{Conclusions}

We study the scattering and transport properties of the Power-law
Banded Random Matrix (PBRM) model and of the three-dimensional (3D)
Anderson model, both at criticality.

We observed a smooth crossover from localized- to
delocalized-like (ballistic-like) behavior in the scattering
properties of the PBRM model by varying $b$ from small ($b\ll 1$)
to large ($b > 1$) values. For this crossover region we proposed
heuristic analytical
expressions for $\langle |S_{aa'}|^2 \rangle_{b}$ and
$\langle |S_{aa'}|^2 \rangle_{D_2}$.
For small $b$, we have shown that $\rho(\ln T)$ is scale invariant
with the typical value of $T$, $T_{\tbox{typ}}$, as scaling
factor.
We realized that RMT results, expected in the limit $b\to \infty$,
are already recovered for relatively small values of the bandwidth:
$b\ge 10$. However, this fact is closely related to the small
number of attached leads we used in this work: the larger the
number of leads the larger the value of $b$ needed to approach
RMT behavior.

Our conclusions are valid for leads attached to the bulk
of the system as well as for leads attached at the boundary.
In this work we assumed that time-reversal symmetry ($\beta=1$)
is present in our disordered systems.
Moreover, the case of broken time-reversal symmetry ($\beta=2$)
has been preliminarily studied in Ref.~[\onlinecite{AMV09}] where
similar conclusions to this work were made.

We have also shown that the scattering properties of the 3D
Anderson model are similar to those for the PBRM model with
$b \in [0.2,0.4]$. This result is in agreement with previous
studies~\cite{EM00b} were it was shown that several
critical quantities related to the spectrum and eigenstates
of the PBRM model with $b=0.3$ are practically the same as
for the 3D Anderson model. This makes the PBRM model an
excellent candidate to explore the properties of the 3D
Anderson model at criticality at a low computational cost.

We want to stress that for both models at criticality,
the PBRM model and the 3D Anderson model, we found the
universal behavior $\rho(\ln T) \propto T^{M^2/2}$ for $T\ll 1$,
that we tested here for $M=1$ and $M=2$.
Moreover by the use of the PBRM model
with $\beta=2$ (not shown here) we have already verified
that the more general expression
$\rho(\ln T) \propto T^{\beta M^2/2}$,
derived from~\cite{MK04}
$w(T) \propto T^{\beta M^2/2-1}$
with $0<T<1$, holds.

We emphasize that, even though we used a scattering setup
where the leads are attached perfectly to sites at one side 
of the sample, our conclusions are not restricted to this 
topology.\cite{unpublished}

Finally, we recall that we concentrate here on the case of
a small number of single-channel attached leads (up to ten).
The study of the scattering and transport properties of the
PBRM and the 3D Anderson models in the regime of $M\gg 1$
will be the subject of a forthcoming
contribution.\footnote{In Ref.~[\onlinecite{MG09b}] some
properties of the transmission for the PBRM model were
already studied for $M\gg 1$.}

\begin{acknowledgments}
We thank one of the Referees for valuable comments.
This work was partially supported by the Hungarian-Mexican
Intergovernmental S \& T Cooperation Program under grants
MX-16/2007 (NKTH) and I0110/127/08 (CONACyT). We also
acknowledge support form CONACyT Mexico grant CB-2006-01-60879;
the European Social Fund; The MICINN (Spain) under Project
No. FIS 2009-16450 and through the Ram\'on y Cajal Program;
the Alexander von Humboldt Foundation; and the Hungarian
Research Fund (OTKA) grants K73361 and K75529.
\end{acknowledgments}

\end{document}